\documentclass[twocolumn,superscriptaddress,showpacs,preprintnumbers,amsmath,amssymb,aps]{revtex4}

\usepackage{graphicx}
\usepackage{dcolumn}
\usepackage{bm}
\usepackage{longtable}
\begin{document}

\title{Enhanced thermodynamic efficiency in time asymmetric ratchets}
\vskip.6cm
\author{Raishma Krishnan}
\email{raishma@iopb.res.in}
\author{Soumen Roy}
\email{sroy@iopb.res.in}
\author { A. M. Jayannavar }
\email{jayan@iopb.res.in}
\affiliation{Institute of Physics, Sachivalaya Marg, 
Bhubaneswar 751005, India }

\begin{abstract}
Abstract: The energetic  efficiency of an overdamped 
 Brownian particle in a saw tooth potential  in the  presence of time
asymmetric forcing is studied in the adiabatic limit.  An error 
made in the earlier work on the same problem in literature is 
corrected. We find that asymmetry in potential 
together with temporal asymmetry in forcing leads to much enhanced 
efficiency without fine tuning of parameters. The origin of this is 
traced to the suppression of backward current.  
We also present a comparative study between the role of continuous and
discontinuous ratchet forces on these measurable quantities. 
We find that the thermal fluctuations can optimize the energy transduction, the
range of parameters, however,  being very small. This ratchet model 
also displays current reversals on tuning of 
parameters even in the adiabatic regime. The possible relationships 
between nature of currents, entropy production and input energy 
are also addressed.\\

\end{abstract}

\maketitle
 \section{Introduction}

The study of the nature of directed motion induced by 
random noise in periodic systems in the absence of a bias has attracted wide
interest. By now, the rectification of thermal fluctuations have become a major
area of research in nonequilibrium statistical mechanics. The presence 
of unbiased nonequilibrium perturbations, either stochastic or 
deterministic, together 
with a broken spatial or temporal asymmetry play a key role in obtaining 
directed motion without violating the second law of thermodynamics. 
Such systems or ratchets convert nonequilibrium 
fluctuations into useful work in the presence of load. Moreover, in these
systems, noise play a constructive role (i.e., transformation of noise in
spatially periodic systems into directed current). A large family of models 
of Brownian ratchets~\cite{julicher,reiman,1amj,special} have 
been introduced to obtain insight into the basic
mechanism of noise rectification. Some of them are 
flashing ratchets, rocking ratchets, time asymmetric ratchets, frictional
ratchets etc~\cite{reiman}. Numerous studies have been carried 
out to understand 
the nature of currents, their possible reversals and also the 
efficiency of energy transduction. The results obtained are utilized to 
develop proper models that efficiently separate particles of micro and nano
sizes and also for the development of machines at nano scales~\cite{special}. 
Such models are also the prototype to understand the basic 
mechanism of operation of molecular
motors or protein molecules in our cells that transfer cargo and 
organelles very efficiently in a very noisy environment. It  
also has extensions in game theory under the name of 
Parrondo's paradox~\cite{web}. These are basically 
counter intuitive games based on translation of the dynamics 
of Brownian particle in a flashing ratchet to
gambling games. Here, two losing games (or strategies), when 
alternated randomly or periodically give rise to a winning game. 
These paradoxes have a profound role in several multidisciplinary areas.      

With the emergence of a separate subfield called stochastic energetics 
~\cite{sekimoto,parrondo} it is possible to establish the 
compatibility between Langevin or Fokker-Planck formalism, which describes
stochastic dynamics,  and the laws of thermodymanics.  
Using this  framework one can calculate various physical quantities 
such as thermodynamic efficiency of energy transduction~\cite{kamgawa}, 
energy dissipation (hysteresis loss), entropy (entropy production)
~\cite{rkamj} etc., thereby providing a new tool to study systems 
far from equilibrium. 
  
The intrinsic irreversibility associated with ratchet operation makes the
ratchet to be less efficient. For example, the attained value of efficiency 
in flashing and rocking ratchet were found to be 
below the subpercentage regime. However, it
has been shown that at very low temperatures fine tuning of parameters could
lead to a larger efficiency, the regime of parameters being 
very narrow~\cite{sokolov}. Optimization of 
energetic efficiency of the sawtooth ratchet in presence of 
spatial symmetry but in presence of time symmetric
rocking has been worked out in detail in~\cite{sokolov}. Moreover, protocols to
optimize the efficiency is given in~\cite{sokolov, hernandez}.

Recently Makhnovskii et al.~\cite{tsong} constructed a special type 
of flashing ratchet with two asymmetric double-well periodic-potential-
states displaced by half a period. Such flashing ratchet 
models were found to be 
highly efficient with efficiency an order of magnitude higher 
than in earlier models~\cite{sekimoto,parrondo,kamgawa,astu05}. The basic 
idea behind this enhanced efficiency is that even for 
diffusive Brownian motion the choice of appropriate potential 
profile ensures suppression of backward motion and hence reduction in the 
accompanying dissipation. Similar to the case of flashing ratchets
~\cite{tsong}, we had earlier \cite{pre} studied the motion of 
a particle in a rocking ratchet  by applying a temporally asymmetric 
but unbiased periodic forcings~\cite{ajdari,chialvo,phylett,ai} in 
the presence of a sinusoidal potential. The efficiency obtained was 
very high, much above the subpercentage level (about $\sim 30-40 \%$ 
without fine tuning) in the presence of temporal asymmetry 
alone. 

In the present work we study the same problem but in a saw tooth
potential and make a comparison so as to elucidate the sensitivity of these
physical quantities on the smoothness or regularity of the underlying ratchet
potential. The important underlying  factor  is the temporal 
asymmetry~\cite{ajdari,chialvo,phylett,ai} in the external forcing 
which leads to noise induced currents in the absence of external bias even 
for the case of spatially symmetric potential. In this adiabatically 
rocked time asymmetric correlation ratchet, a larger force field 
is applied for a short time interval of period in one 
direction as compared to a smaller force for a longer time interval in
the other direction, see Fig.~\ref{fig-sawtooth}. Some qualitative 
differences between the 
smooth and piecewise linear ratchet potential which are observed 
is discussed. The surprisingly sensitive dependence of the physical 
quantities such as unidirectional current on the degree of regularity 
or smoothness of the ratchets (continuous and discontinuous 
forces) has been demonstrated by Doering et al~\cite{doering}. 

Ai et al.~\cite{ai} have also studied the same problem of a Brownian
particle moving in a periodic saw tooth potential subjected to a temporally
asymmetric periodic rocking. However, there is an error in the expression 
for the energy per unit time that a ratchet gets from the 
external force or in other words, the input energy~\cite{pre}. 
In this work we take into account this correction 
and have calculated the efficiency and other physical quantities 
and presented our results. We find that the temporal asymmetry in driving 
enhances the efficiency in a very significant manner even 
for a spatially symmetric potential. Also, in the presence of spatial 
asymmetry in potential, the efficiency is found to be 
almost $90\%$ at low temperature.  Current reversals are also observed 
in the parameter space of operation even in the adiabatic regime. 

We also present our analysis of the behaviour of entropy production, 
current and input energy with temperature in this ratchet system.  
In the absence of any bias the noise induced currents show a 
peak with temperature.  The question that naturally arises is 
whether this peak is related to underlying resonance (stochastic resonance 
~\cite{gammaitoni-rmp}) due to the synchronization of the position of the 
particle with the external drive induced by the noise. 
Our analysis of input energy $E_{in}$, rules out the presence 
of any resonance features in the dynamics 
of the position of the particle in these systems in 
the adiabatic regime~\cite{rkamj,entropy-rk}. This follows 
from the earlier works which show that the
existence of stochastic resonance in the dynamics of the 
particle is revealed by
a peak in the input energy~\cite{iwai,stoc-bonofide}.

The onset of unidirectional currents in ratchet systems can also be 
viewed as an example of temporal order coming out of disorder. This can happen
only at the expense of an overall increase in the entropy production 
in the system along with its environment. Thus one expects 
a correlation between the maxima in current and the maxima in 
entropy production. However, our results show that the maxima in current and 
entropy production do not correlate with each other.\\
 
 \section{The Model:}
A simple model for our ratchet system is described by the stochastic
differential equation (Langevin equation) for a Brownian particle in the
overdamped regime. This is given by~\cite{overdamp}

\begin{equation}
 {\dot{q}} = {- {\frac{V^\prime(q)-F(t)+L}{\gamma}}}
 +  \xi(t),\label{lang}
 \end{equation}
where $\xi(t)$ is a randomly fluctuating Gaussian 
thermal noise with zero mean and correlation, 
$<\xi(t)\xi(t^\prime)>\,=\,(2\,{k_BT}/\gamma) \delta(t-t^\prime)$. 
 
\begin{figure}[htp!]
 \begin{center}
\input{epsf}
\includegraphics [width=2.8in,height=1.9in] {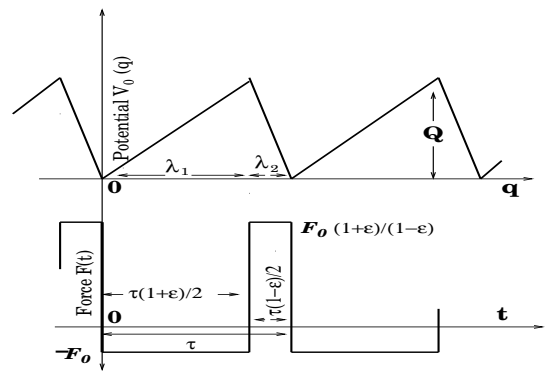}
\caption{Schematic representation of the potential and the external force 
$F(t)$ as a function of space and time respectively. } 
\label{fig-sawtooth}
 \end{center}
 \end{figure}

In the present work we consider the piecewise linear ratchet potential,
Fig.~\ref{fig-sawtooth},  as in 
the case of Magnasco et al.~\cite{magnasco} with periodicity $\lambda$ set equal
to unity. This also corresponds to the spacing between the wells. 
We later on scale all the lengths with respect to $\lambda$. 
$F(t)$ which corresponds to the externally applied 
time asymmetric force with zero average over the period is also shown in
Fig.~\ref{fig-sawtooth}.  The force in the gentler and
steeper side of the potential are respectively $f^+=\frac{-Q}{\lambda_1}$  
and $f^-= \frac{Q}{\lambda_2}$ and $Q$ is the height of the potential. 
In the above expression we have also included
the presence of an external load $L$, which is essential for defining
thermodynamic efficiency.   
Following Stratonovich's interpretation~\cite{stratanovich}, 
the corresponding Fokker-Planck equation~\cite{riskin} is given by
\begin{eqnarray}
\frac {\partial P(q,t)}{\partial t}&=& \frac {\partial}{\partial q}
\Big[k_BT \frac{\partial P(q,t)}{\partial q}\\ \nonumber
&+& [V^\prime(q)-F(t)+L]P(q,t)\Big].\label{fk}
\end{eqnarray} 
Since we are interested in 
the adiabatic limit we first obtain an expression 
for the probability current density $j$ in the presence of a 
constant external force $F$. The expression for current~\cite{magnasco} is
\begin{widetext}
\begin{equation}
j(F)= \frac{{P_2}^2 \sinh\{\lambda[F-L]/2k_BT\}}{{{k_BT{(\lambda/Q)}^2}P_3} -
{(\lambda/Q)P_1P_2 \sinh \{\lambda[F-L]/2k_BT\}}}
\end{equation}
\end{widetext}
where 
\begin{eqnarray}
P_1 &=& \Delta + \frac{\lambda^2-\Delta^2}{4} \frac{F-L}{Q}\\
P_2 &=& (1-\frac{\Delta [F-L]}{2Q})^2-(\frac{\lambda[F-L]}{2Q})^2\\
P_3 &=& \cosh(\{Q-0.5\Delta[F-L]\}/k_BT)- \nonumber \\
    &  & \mbox{}  \cosh\{\lambda[F-L]/2k_BT\}
\end{eqnarray}

where $\lambda=\lambda_1+\lambda_2$ and $\Delta =\lambda_1 - \lambda_2$, 
the spatial asymmetry parameter. The current in the 
stationary regime averaged over the
period $\tau$ of the driving force $F(t)$  is given by 
\begin{equation}
<j>=\frac{1}{\tau}\, \int_0^\tau \, j(F(t))\,dt.
\end{equation}    
We assume that $F(t)$ changes slow enough, 
i.e., its frequency is smaller than any other 
frequency related to the relaxation rate 
in the problem such that the system is in a steady state at each 
instant of time. 

In the present work we consider time asymmetric ratchets 
with a zero mean periodic driving force~\cite{pre,chialvo,ai} given by
\begin{eqnarray}
F(t)&=& \frac{1+\epsilon}{1-\epsilon}\, F_0,\,\, \{n\tau 
\leq t < n\tau+ \frac{1}{2} \tau (1-\epsilon)\}, \\ \nonumber
    &=& -F_0,\,\, \{n\tau+\frac{1}{2} \tau(1-\epsilon) < t \leq (n+1)\tau\}.
\end{eqnarray}
Here, the parameter $\epsilon$ signifies the temporal asymmetry 
in the periodic forcing, $\tau$ the period of the driving force $F(t)$ 
and $n=0,1,2....$ is an integer. For this forcing in the adiabatic limit the 
expression for time averaged current is~\cite{chialvo,kamgawa}
\begin{eqnarray}
<j>=j^+ + j^-\,,
\end{eqnarray}
with 
\begin{eqnarray}
j^+&=& \frac{1}{2}(1-\epsilon)\, j(\frac{1+\epsilon}{1-\epsilon}F_0)\,,\\ \nonumber \label{jplus}
j^- &=& \frac{1}{2}(1+\epsilon)\,j(-F_0) \label{jminus} 
\end{eqnarray}
where $j^+$ is the 
current fraction in the positive direction over a fraction of time period 
$(1-\epsilon)/2$ of $\tau$ when the external driving force field is 
$(\frac{1+\epsilon}{1-\epsilon})F_0$ and $j^-$ is 
the current fraction over the  time period 
$(1+\epsilon)/2$ of $\tau$ when the external driving force 
field is $-F_0$. 
The input energy $E_{in}$ per unit time is given by~\cite{kamgawa,pre} 
\begin{eqnarray}
E_{in}=F_0 [(\frac{1+\epsilon}{1-\epsilon})j^+-j^-].\label{Ein}
\end{eqnarray}  
In order that the system does useful work a load $L$ is applied in a direction 
opposite to the direction of current in the ratchet. The overall 
potential is then  $V(q)=[V_0(q) + q L]$. 
As long as the load is less than the stopping force $L_s$ current flows against
the load and the ratchet does work.  Beyond the stopping force 
the current flows in the same direction as the load and hence no 
useful work is done. Thus in the operating range of the load, 
$0<\,L<\,L_s$, the Brownian particles move in the direction 
opposite to the load and the ratchet does useful work (storing energy in the
form of potential or say, charging the battery). 
The average work done over a period is given by~\cite{kamgawa} 
\begin{eqnarray}
E_{out}= L [j^+ + j^-]\,.\label{Eout}
\end{eqnarray}
The thermodynamic efficiency of energy transduction is 
~\cite{sekimoto,parrondo} 
\begin{equation}
\eta=  \frac{L [j^+ + j^-]}{F_0 [(\frac{1+\epsilon}{1-\epsilon})j^+-j^-]}. \label{eta}
\end{equation}  

In the limit when the current fraction in the forward direction is much larger
than that in the backward direction, $j^+ >> j^-$, and at very low temperature 
(temperature tending to zero) the efficiency is given by~\cite{pre} as
\begin{equation}
\eta = \frac{L(1-\epsilon)}{F_0(1+\epsilon)} \label{limiteta}.
\end{equation}

The suppression of backward current at low temperature occurs for values of
$F_0$ less than $Q/\lambda_2$. However, finite current fraction flows in the
positive direction when $\frac{(1+\epsilon)}{(1-\epsilon)}F_0 >
-\frac{Q}{\lambda_1}$ or $F_0 > \frac{Q(1-\epsilon)}{\lambda_1(1+\epsilon)}$. 
Hence, in the operating range of $F_0$, $\frac{Q}{\lambda_2} > F_0 >
\frac{Q(1-\epsilon)}{\lambda_1(1+\epsilon)}$, a high efficiency is expected in
the low temperature regime~\cite{sokolov}.

In the absence of a load the particle moves in a periodic potential without 
tilt and hence the system does not store any energy. Consequently all the input 
energy in the steady state is dissipated away. In such a case 
the energy loss in the medium $E_L=E_{in}$. 
$E_L$ inturn is equal to the heat $Q_h$ transferred to the bath and 
thus entropy production $S_p=Q_h/T=E_L/T$~\cite{parrondo}. 
Thus the total increase in the entropy  (or the entropy production) 
of the bath (universe) integrated over the period of the external drive 
is given by~\cite{parrondo} 
$$S_p = \frac{Q_h}{T}=\frac{E_{in}}{T}=\frac{E_L}{T}.$$  
As discussed in the introduction, 
currents in ratchet systems are generated at the expense of entropy 
and thus we expect a correlation between the magnitude of current 
and the total entropy production.

In our work all the physical quantities are taken in 
dimensionless units. Moreover, the energies and lengths are scaled with respect
to $Q$, the barrier height and $\lambda$, the spatial period of the potential 
respectively. In the following section we  present our results and 
the discussion of our calculations.

 \section{Results and Discussions}

We study the motion of an overdamped Brownian particle subjected 
to a time asymmetric periodic forcing but in presence 
of a saw tooth potential. We present the noticeable differences 
between the motion in a smooth potential as in~\cite{pre} with 
that in a piecewise linear saw tooth potential. The role of 
smoothness or regularity in potential on the efficiency of 
energy transduction is clearly presented here.

\begin{figure}[hbp!]
 \begin{center}
\input{epsf}
\includegraphics [width=2.8in,height=1.9in] {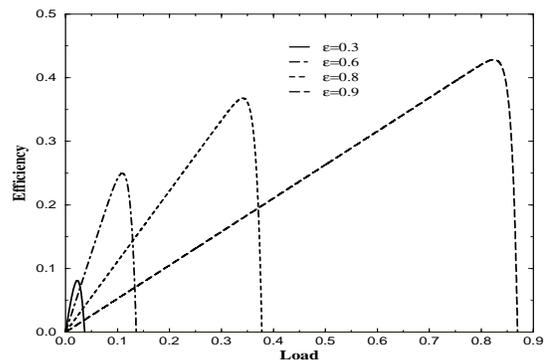}
\caption{Efficiency vs load for  $\Delta=0.0,\, F_0 =0.1, \, T=0.01,\, 
Q=1$ with varying  $\epsilon$.}\label{eff_Ldel0}
 \end{center}
 \end{figure}
 
To start with, in Fig.~\ref{eff_Ldel0} we study the behaviour 
of efficiency with load in a spatially {\textit{symmetric}} saw tooth 
potential $(\Delta=0)$ in the presence  
of time asymmetric driving field for fixed values of $F_0=0.1$, $T=0.01$ 
and $Q=1$ for different values of $\epsilon$. Currents in this 
ratchet model arise solely due to the temporal asymmetry 
factor.  For a given $\epsilon$, the
efficiency increases as a function of load and then decreases. 
The attained value of efficiency is much higher than those attained
in other models and it  keeps increasing with increasing
$\epsilon$. The stopping force $L_s$ too is found to increase with increase 
in $\epsilon$. Large the $\epsilon$, larger will be the current and efficiency 
as long as $F_0$ is less than the critical field, so that the barriers to motion
in one direction alone disappears and there will be no current in the
opposite direction. We  notice that efficiency depends linearly on the load as
long as $L$ is much less than $L_s$ where the backward motion is suppressed and
the slope is given by $\frac{(1-\epsilon)}{(1+\epsilon)}F_0$ consistent with
Eqn.~\ref{limiteta}. In contrast to the case of smooth
sinusoidal potential~\cite{pre} the locus of the
peak in efficiency monotonously increases in the saw tooth case. 
The value of efficiency is also much higher than that obtained in the 
smooth potential case. The input energy, $E_{in}$, output energy, $E_{out}$, the
fraction of currents $j^{+}$, $j^-$ and the average current $<j>$ 
show the same qualitative behaviour as a function of load as is seen 
in Fig.3 of~\cite{pre}  and the observed behaviour 
has been discussed in detail in reference~\cite{pre}. Hence we do not deal with
these quantities separately in the present work.

  \begin{figure}[hbp!]
 \begin{center}
\input{epsf}
\includegraphics [width=2.8in,height=1.9in] {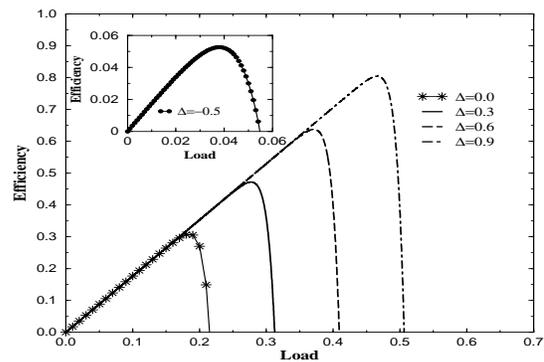}
\caption{Efficiency vs load for  various $\Delta=0.9,\,0.6,\,0.3,\,0.0$ 
with fixed
$ F_0 =0.1$ $\epsilon=0.7$, $T=0.01$ and  $Q=1$. Inset shows the efficiency 
for $\Delta=-0.5$ with other parameters remaining the same. } 
\label{eff_Lvarydel}
 \end{center}
 \end{figure}

As noted in Fig.~\ref{eff_Ldel0} one can attain an efficiency 
of the order of $40\%$ for given physical parameters 
for the spatially symmetric $(\Delta=0.0)$ rocked ratchet. We now explore the
additional role of spatial asymmetry on the above results. For that, in 
Fig.~\ref{eff_Lvarydel} we plot efficiency 
as a function of load for various asymmetry in potential $(\Delta)$ with fixed 
$F_0=0.1$, $T=0.01$, $Q=1$ and $\epsilon=0.7$. 
We observe that an asymmetry in the potential enhances efficiency
and also increases the range of operation of the ratchet. As in the smooth
potential case, the higher the $\epsilon$, the larger the current 
and hence a larger load is necessary for the current 
to reverse its direction. From this figure it is clear that 
we can obtain a peak value of efficiency of the order of $30 \%$ even 
in the absence of spatial asymmetry. This peak value of efficiency 
and the range of operation of the load increases for higher asymmetry. 
For $\Delta=0.9$ we obtain a peak value of
efficiency of more than $80 \%$ which is very high given the fact that ratchet
operates in an irreversible mode. It should also be noted that the 
initial slope of the efficiency versus load curve (for $L<L_s$) is the same, 
i.e., independent of $\Delta$, again in consistency with Eqn.~\ref{limiteta}. 
We can conclude from the above figure that additional spatial asymmetry 
will further help in enhancing the efficiency of time asymmetric ratchets. 
This is also due to the fact that currents due to the finiteness in the
spatial asymmetry parameter $\Delta$ and it being positive enhances 
the currents in the system as compared to the case when $\Delta=0.0$. 
Opposite conclusions will be reached on the effect of
$\Delta$ on efficiency if $\Delta$ is negative, which is obvious. 
The reduction in currents when $\Delta$ is negative and $\epsilon$ is positive
will be discussed in detail later in connection with current reversals. 

In the inset of Fig.~\ref{eff_Lvarydel} we plot efficiency as a function 
of load for a representative positive and negative value of $\epsilon$ and
$\Delta$ respectively. Here, one can clearly notice that the attained 
efficiency is in the subpercentage regime. Our further analysis will be
restricted to the case wherein $\Delta$ and $\epsilon$ remains 
positive as in this parameter space  we naturally  expect high efficiency.  
\begin{figure}[hbp!]
 \begin{center}
\input{epsf}
\includegraphics [width=2.8in,height=1.9in] {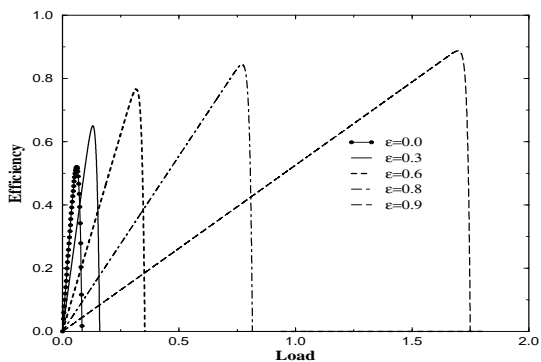}
\caption{Efficiency vs load for  $\Delta=0.9,\, F_0 =0.1, \, T=0.01,\, 
Q=1$ with varying  $\epsilon$.}\label{eff_Ldel9}
 \end{center}
 \end{figure}

We now study the role of the temporal asymmetry parameter 
$\epsilon$ for the case 
of spatially asymmetric $(\Delta=0.9)$ ratchets. Fig.~\ref{eff_Ldel9} 
shows the behaviour of efficiency as a function of load for varying
$\epsilon$ for fixed value of $\Delta=0.9$. It is clear that the inclusion of
time asymmetry leads to enhanced value of efficiency and the operational 
range of load. An efficiency of about 
$\sim 90 \%$ is readily attained as can be seen in Fig.~\ref{eff_Ldel9}. 
The locus of the peak value in efficiency monotonously increases with increase 
in $\epsilon$. This is in contrast to the non monotonic behaviour observed in a 
smooth sinusoidal potential~\cite{pre}. Moreover, the efficiencies are much
higher for these ratchets with discontinuous potential. For the case
$\epsilon=0$ we get an efficiency of $\sim 40 \%$. Such a case with $\epsilon=0$
and finite $\Delta$ is discussed in ~\cite{sokolov}.
 As has been mentioned earlier, the initial slopes are linear 
in accordance with Eqn.~\ref{limiteta}. There are some studies in the
deterministic limit where one can attain efficiency to the ideal limit
($\eta=1$). However, these ratchets work in a reversible quasi-static mode of
operation but not in the adiabatic regime~\cite{parrondo,sokolov}. The protocols
of the operation rely on synchronizing the dynamics of the particle with
external force~\cite{parrondo,sokolov}. 

\begin{figure}[htp!]
 \begin{center}
\input{epsf}
\includegraphics [width=2.8in,height=1.9in] {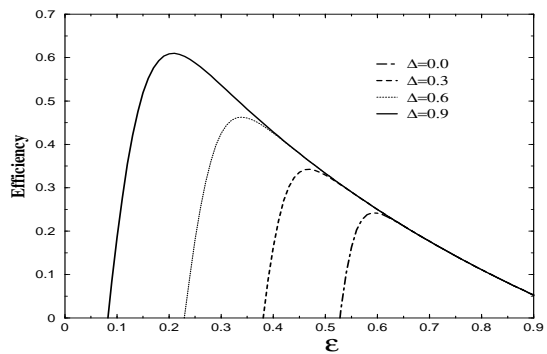}
\caption{Efficiency vs $\epsilon$ for various values of $\Delta$ 
with fixed $F_0 =0.1$, $Q=1$, $L=0.1$ and $T=0.01$.} \label{eff_eps}
 \end{center}
 \end{figure}
 
In Fig.~\ref{eff_eps} we plot the efficiency as a function of $\epsilon$ for
different strength of potential asymmetry for $F_0=0.1$, $L=0.1$, $Q=1$ and
$T=0.01$. Similar to the earlier figure we see that the potential asymmetry
increases the efficiency value. Larger the asymmetry in potential lower is the
value of $\epsilon$ for which one gets higher efficiency. This follows from the
fact that larger the $\Delta$, the smaller is the critical value of $\epsilon$ 
to get current in the forward direction. The critical value of $\epsilon$, 
$\epsilon_c$,  in the absence of load is given 
by $\epsilon_c = \frac{Q_0 -F_0\lambda}
{Q_0 +F_0\lambda}$. One can notice that this critical value decreases as $F_0$
increases. In the absence of load the current vanishes for $\epsilon=0.0$ and
moreover the current fraction in the positive 
direction $j^+$ vanishes as $\epsilon
\rightarrow 1$. Hence naturally a peak is expected in  efficiency as a
function of $\epsilon$. For higher values of $\epsilon$ in the regime where the
backward current is suppressed the slope in the figure is consistent with 
Eqn.~\ref{limiteta} (which is again independent of $\Delta$ as clearly seen in
the figure).
\begin{figure}[htp!]
 \begin{center}
\input{epsf}
\includegraphics [width=2.8in,height=1.9in] {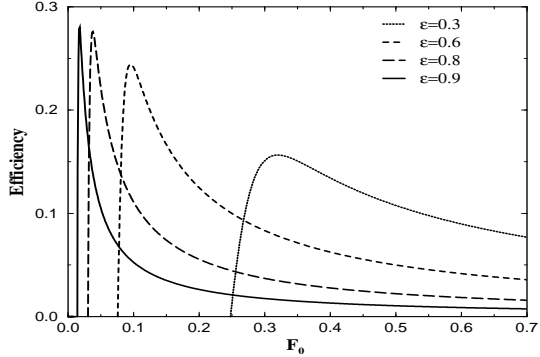}
\caption{Efficiency vs $F_0$ for various values of $\epsilon$ for the symmetric
case with fixed $L =0.1$, $Q=1$ and $T=0.01$.} \label{eff_F0-del0}
 \end{center}
 \end{figure}

\begin{figure}[hbp!]
 \begin{center}
\input{epsf}
\includegraphics [width=2.8in,height=1.9in] {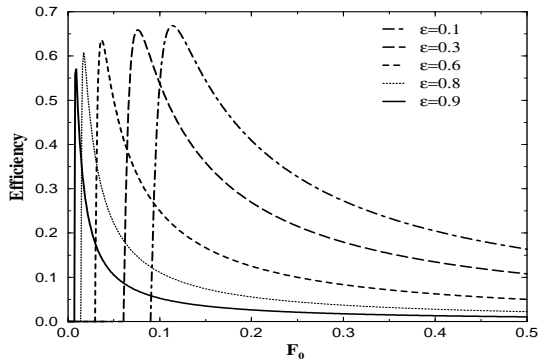}
\caption{Efficiency vs $F_0$ for various values of $\epsilon$ 
with fixed $L=0.1$, $Q=1$, $\Delta=1$ and $T=0.01$.} \label{eff_F0-del1}
 \end{center}
 \end{figure}

In Fig.~\ref{eff_F0-del0} we plot efficiency as a function of $F_0$ 
for the case of symmetric potential $(\Delta=0.0)$ for different 
values of $\epsilon$ with fixed 
$L=0.1$, $Q=1$ and $T=0.01$. Consistent with the general observation of this
problem, for lower $\epsilon$ values we need larger $F_0$ to get 
forward current. Moreover, in the absence of load, the current 
vanishes in both zero $F_0$ and large $F_0$ limit. 
In the large $F_0$ limit, the barriers to motion in 
the forward as well as backward direction disappear and consequently the 
average current over the period vanishes. Thus a peak in the efficiency as 
a function of $F_0$ is obvious. Additional spatial asymmetry enhances 
the efficiency by a large amount. This can be clearly seen in 
Fig.~\ref{eff_F0-del1} where we have plotted efficiency versus 
$F_0$ for the case $\Delta=1.0$. The difference between 
Figs.~\ref{eff_F0-del0} and ~\ref{eff_F0-del1} is that the 
envelope of the peak value of efficiency show opposite behaviour. 
In the case of smooth potential we had observed earlier~\cite{pre} 
that the envelope of the peak of efficiency decreases 
with increase in $\epsilon$ in contrast with that in Fig.~\ref{eff_F0-del0}.
\begin{figure}[htp!]
 \begin{center}
\input{epsf}
\includegraphics [width=2.8in,height=1.9in]{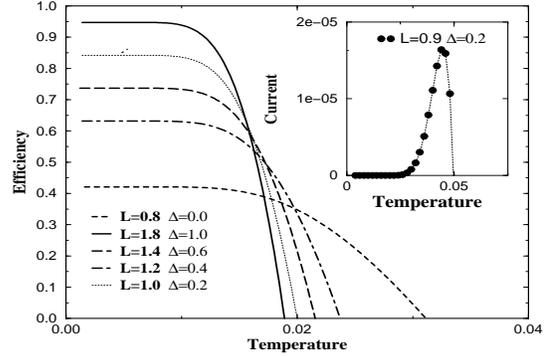}
\caption{Efficiency vs temperature for various values of load and $\Delta$ 
with fixed $Q=1$, $F_0=0.1$ and $\epsilon=0.9$. Inset shows 
the peaking of current with temperature for $\Delta=0.2$, $F_0=0.1$, $L=0.9$, 
$Q=1.0$ and $\epsilon=0.9$.} 
\label{eff_Tinsetcurrent} 
 \end{center}
 \end{figure}

\begin{figure}[hbp!]
 \begin{center}
\input{epsf}
\includegraphics [width=2.8in,height=1.9in]{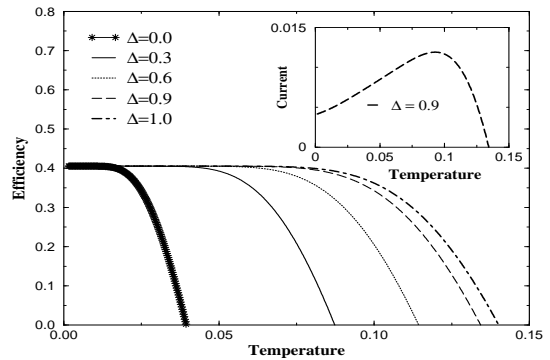}
\caption{Efficiency vs temperature for various values of $\Delta$ 
with fixed $\epsilon=0.9$, $F_0=0.1$, $L=0.77$ and $Q=1$.} 
\label{eff_Tvarydel}
 \end{center}
 \end{figure}
So far we have shown that a large efficiency of the 
order of unity can be obtained
readily in the time asymmetric rocked ratchets in presence of additional spatial
asymmetry. Notably, this large efficiency is obtained in the irreversible mode
of operation in the adiabatic regime. In presence of both $\Delta$ and
$\epsilon$ we do not have to fine tune the parameters and 
we get much higher efficiency above the subpercentage limit. 
In the following we address the question, can thermal 
fluctuations (noise) facilitate energy transduction?, a subject which has been
pursued widely and is of fundamental importance in its own right in the areas in
which noise play a constructive role~\cite{kamgawa}.  

In Fig.~\ref{eff_Tinsetcurrent} we plot the efficiency as a function of 
temperature for various load and $\Delta$ with fixed
$\epsilon=0.9$. We observe that the efficiency decreases with noise strength 
$(T)$. We find the value of efficiency at very low temperature 
to be exactly coinciding with the values obtained from the 
analytical expression for efficiency in the limit
$j^+ >> j^-$, Eqn.\ref{limiteta}. 
\begin{figure}[htp!]
 \begin{center}
\input{epsf}
\includegraphics [width=2.8in,height=1.9in] {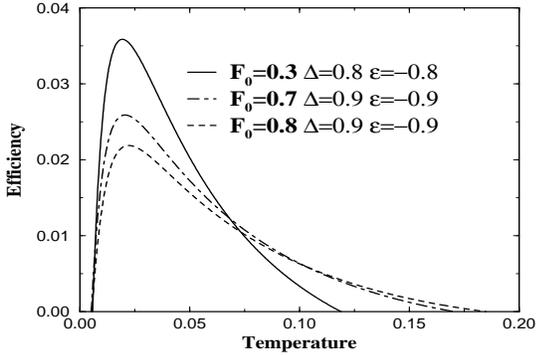}
\caption{Efficiency vs temperature for  three different cases of physical
parameters. (i) $F_0=0.3$, $\Delta=0.8$, $\epsilon=-0.8 $ 
(ii)$F_0=0.7$, $\Delta=0.9$, $\epsilon=-0.9$ (iii) $F_0=0.8$, 
$\Delta=0.9$, $\epsilon=-0.9$  for fixed  $ L=0.01$, 
and $Q=1$.} \label{eff_Tpeak}
 \end{center}
 \end{figure}

In Fig.~\ref{eff_Tvarydel} we plot efficiency as a function of temperature 
for different spatial asymmetry parameter $\Delta$ with fixed $L=0.77$, 
$\epsilon=0.9$, $Q=1$ and $F_0=0.1$. One can notice readily that at low
temperatures efficiency is independent of $\Delta$, Eqn.\ref{limiteta}, 
and it decreases with temperature. Also, as 
one increases $\Delta$ a larger range of temperature 
is obtained over which the efficiency value is high. 
In the parameter range we have considered we generally 
observe that temperature (noise) cannot facilitate energy transduction i.e., 
it cannot optimize the efficiency. This is in spite of the fact that in all the
cases current as a function of temperature exhibits a peaking behaviour ( for
example see the inset of Figs.~\ref{eff_Tinsetcurrent} and 
~\ref{eff_Tvarydel})  for a representative parameter value mentioned 
in the figure captions.
\begin{figure}[hbp!]
 \begin{center}
\input{epsf}
\includegraphics [width=2.8in,height=1.9in] {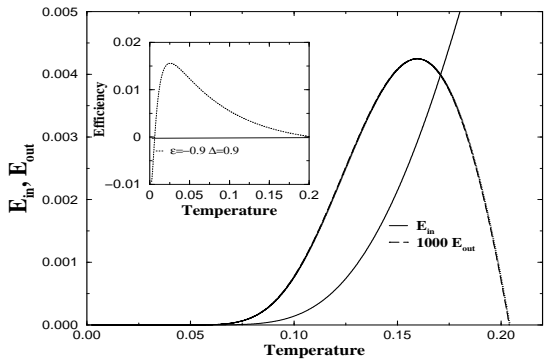}
\caption{Efficiency vs temperature for  $\Delta=0.9,\, F_0 =0.10, \, L=0.01,\, 
Q=1$ and $\epsilon=-0.9$. The inset shows the behaviour of input and output
energy for the same set of parameters. The output energy curve is blown up by a
factor of $1000$ to make it consistent with the scale chosen.}
\label{eff_Tinoute-9}
 \end{center}
 \end{figure}

However, with a judicial choice of parameters which require fine tuning we
obtain a regime in parameter space where the efficiency exhibits a peak with
temperature. In this parameter range, temperature or noise facilitates energy
transduction. Fig.~\ref{eff_Tpeak} shows the peaking behaviour of 
thermodynamic efficiency with temperature for three representative 
sets of parameters mentioned in the figure caption. 
The magnitude of current and efficiency are, however, quite small in this
range which we have verified separately. 

\begin{figure}[htp!]
 \begin{center}
\input{epsf}
\includegraphics [width=2.8in,height=1.9in] {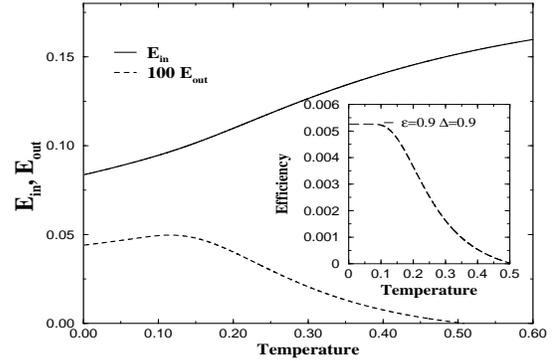}
\caption{Efficiency vs temperature for  $\Delta=0.9,\, F_0 =0.1, \, L=0.01,\, 
Q=1$ and $\epsilon=0.9$. The inset shows the behaviour of input and output
energy for the same set of parameters. The output energy curve is blown up by a
factor of $100$ to make it consistent with the scale chosen.}
\label{eff_Tinoute9}
 \end{center}
 \end{figure}

To understand this behaviour of efficiency with temperature, in 
Fig.~\ref{eff_Tinoute-9} we plot the input energy, $(E_{in})$, and the output
work, $(E_{out})$, as a function of temperature. The input 
energy is found to increase monotonically with temperature. 
However, the output energy shows a peak with temperature. The 
output energy curve is blown up by a factor of $1000$ to make 
it comparable with the scale chosen. At very low temperature 
$(T<0.006)$ the efficiency is negative. The current in the 
absence of load is very small in this regime. For a given 
applied load, the current flows in the direction of the load 
and consequently the output energy is also negative 
(which could not be seen on the scale we have 
chosen in the figure). The output energy then increases with temperature 
and becomes positive for $T>0.06$. At the crossover points the 
finite value of the input energy give rise to zero efficiency 
since the output work is zero. As the temperature is increased the
output work increases non monotonically and then becomes zero 
at a temperature value of about $0.21$, beyond which 
(i.e., beyond the operating range of the load), the current 
flows in the direction of the load. Thus at $T \sim 0.21$ the output energy and
consequently the efficiency is zero. Hence we expect a peaking behaviour in
efficiency as a function of temperature as is shown in the inset 
of the figure. It should be noted that the current in the absence 
of load shows a peak with temperature.

In Fig.~\ref{eff_Tinoute9} we plot the input and output energy for the case
where the efficiency monotonically decreases with temperature. All the physical
parameters are mentioned in the figure captions. In contrast to that observed 
in Fig.~\ref{eff_Tinoute-9}, we note that both the output energy 
and the input energy are finite at zero temperature leading in turn 
to a finite value of efficiency. As we
increase the temperature, the input energy increases monotonically 
whereas the output energy exhibits a small peak. Beyond 
a temperature of $0.52$, $E_{out}$ becomes negative. The rise in 
input energy is very rapid as compared to that of the
output energy and consequently the efficiency decreases monotonically 
with temperature as shown in the inset of the figure up to $T=0.52$ 
beyond which it becomes negative.

So far we have discussed the nature of the efficiency of 
energy transduction as a function of system variables. 
We now concentrate on another aspect in ratchet
systems, namely, current reversals, which play a central role in designing
separation devices. It is known that symmetrically rocked spatially asymmetric
ratchets do not exhibit current reversals in the adiabatic 
regime~\cite{reiman,reiman-reversal,danpremultiple}. However, the
presence of system inhomogeneities (frictional or inhomogeneous ratchets)
can induce single or multiple current reversals even in the 
adiabatic regime~\cite{pre,dan}.
\begin{figure}[htp!]
 \begin{center}
\input{epsf}
\includegraphics [width=2.8in,height=1.9in]{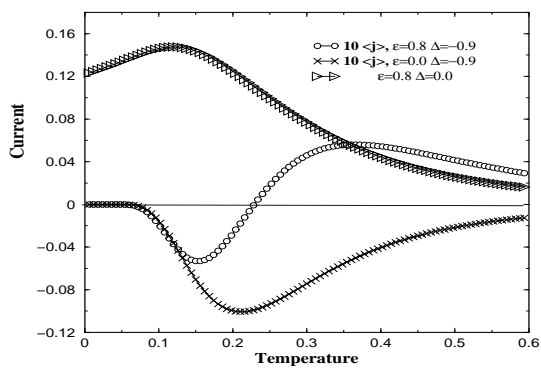}
\caption{Current vs temperature for  $\Delta=-0.9,\, F_0 =0.3,\, 
Q=1$ and $\epsilon=0.8$. $<j>$ is multiplied by a factor of $10$ for the 
cases (i) $\epsilon=0.8$, $\Delta=-0.9$ and 
(ii) $\epsilon=0.0$, $\Delta=-0.9$ to make it
comparable with the scale chosen.} \label{currev2}
 \end{center}
 \end{figure}
In our present case of homogeneous ratchets it is easy to tune current reversals
as there are two asymmetric parameters present in the problem. 
In Fig.~\ref{currev2}  we plot current 
as a function of $T$ for a 
particular value of $\epsilon$ and $\Delta$ given in the caption. 
The parameters are chosen such that the direction 
of current in presence of either of the parameters alone 
should be in opposite directions. For example, in Fig.~\ref{currev2} the
current is in the positive direction when $\epsilon=0.8$ and $\Delta=0$ whereas
it is in the reverse direction when $\epsilon=0.0$ and $\Delta=-0.9$. So
by tuning a combination of these two parameter values for $\epsilon=0.8$ and
$\Delta=-0.9$ one gets current reversal as a function of $T$. 
\begin{figure}[htp!]
 \begin{center}
\input{epsf}
\includegraphics [width=2.8in,height=1.9in] {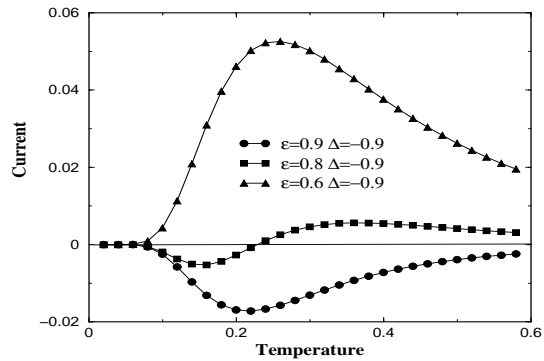}
\caption{Current vs temperature for  $\Delta=-0.9,\, F_0 =0.3,\, 
Q=1$ and varying $\epsilon$.} \label{currev}
 \end{center}
 \end{figure}

It should be noted that this is not an additive effect separately arising from
$\epsilon$ and $\Delta$. The current reversal arises due to complex interplay 
of these two asymmetry parameters. It should be emphasized that once current
inversion upon the variation of one parameter is established, an inversion upon
variation of any other parameter can be readily inferred. For details we refer
to~\cite{reiman}. In accordance with the above reasoning for current reversals,
in Fig.~\ref{currev} we have plotted current versus temperature with fixed value
of $\Delta=-0.9$ and varying $\epsilon$. As we vary $\epsilon$ from large value
to a small value, in the intermediate range of $\epsilon$ we get current
reversal.

Having discussed efficiency and nature of currents and their reversals we now
study other thermodynamic quantities namely, input energy and entropy
production. We would like to find whether any relation exists among them with 
the nature of currents as discussed in the introduction. Some recent studies 
have also tried to reveal the relations between two 
completely unrelated phenomena, namely,  stochastic resonance 
and Brownian ratchets in a formal way through the 
consideration  of  Fokker-Planck equations~\cite{ent-in}. 
Stochastic resonance is a phenomenon
where we can obtain optimal output from a system by adding noise to the system 
~\cite{gammaitoni-rmp}. It has been argued that the rate of 
flow of particles in a Brownian ratchet is analogous to the 
rate of flow of information in the case of stochastic resonance~\cite{allison}. 
\begin{figure}[htp!]
 \begin{center}
\input{epsf}
\includegraphics [width=2.8in,height=1.9in] {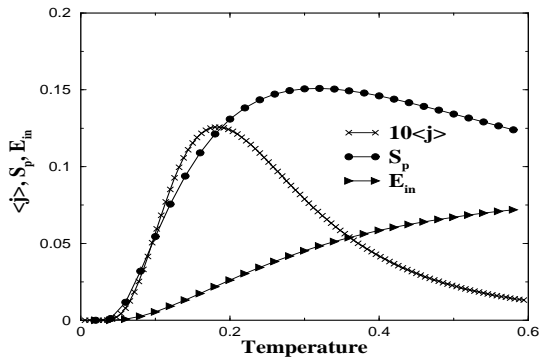}
\caption{$<j>$, $S_p$ and $E_{in}$  vs temperature for  
$\Delta=0.4$ with fixed 
$F_0 =0.1$, $Q=1$ and $\epsilon=0.8$. The current $<j>$ is blown up by a factor
of ten to make it more clearer.} \label{ent-cur-indel4e8}
\end{center}
 \end{figure}

In Fig.~\ref{ent-cur-indel4e8} we plot the entropy production,
current and input energy for a representative case, 
$\Delta=0.4$, $F_0=0.1$ and $\epsilon=0.8$, 
as a function of temperature or noise strength. We observe 
that current exhibits a peak as a function of 
temperature while the input energy is a monotonously increasing function of
temperature~\cite{entropy-rk}. It has been argued earlier that the peak in the
input energy is a good measure for the occurrence of stochastic resonance in the
dynamics of the particle~\cite{iwai,stoc-bonofide}. It is natural to expect at
resonance that the system will extract more input energy from the 
environment which in turn is consequently dissipated away in 
the steady state (for details see ~\cite{iwai,stoc-bonofide}). 
However, the observed monotonic behaviour of the input energy, 
as opposed to the nature of current, rules out the
possibility of any resonance in the dynamics of the particle as a function of
noise strength. 

The presence of net currents in the ratchet increases the 
amount of known information about the system than otherwise. This 
extra bit of information comes from the negentropy or the physical 
information supplied by the external nonequilibrium bath. 
Since the currents are generated at the expense of entropy 
one normally expects  the maxima in current and the maxima in 
the overall entropy production to coincide at the 
same value of noise strength. In fact, in a related development it has been
pointed out that the amount of information transferred by the 
nonequilibrium bath is quantified in terms of algorithmic 
complexity. Moreover, the algorithmic complexity or 
Kolmogorov information entropy exhibits a maxima at the same value 
of physical parameter where the current is maximum~\cite{family}. 
From Fig.~\ref{ent-cur-indel4e8}, we see that the entropy 
production, $S_p$, also exhibits a peak as a function of noise 
strength. The peaks in the average current, $<j>$ 
and total entropy production $S_p$ do not occur at the 
same $T$. This clearly indicates that maxima in the 
entropy production does not  take place at the same 
value where the current is maximum thereby ruling out the 
correlation between the entropy production peak and the peak 
in current maximum~\cite{entropy-rk}.

\section{Conclusions}

We have studied in detail the nature of energetic efficiency driven by zero
average time asymmetric forcing in the adiabatic limit. The potential is taken
to be of the saw tooth type characterized by an asymmetry parameter 
$\Delta$. In the presence of temporal and spatial asymmetry 
we have shown that a much higher efficiency ,above the subpercentage regime 
(known for other ratchets), can be readily obtained. 
Spatial asymmetry together with temporal asymmetry give larger 
efficiency as compared to the presence of spatial or temporal asymmetry alone. 
At low temperatures an efficiency value closer to the ideal limit 
can be obtained by judicious tuning of physical parameters even though 
the operation of the ratchet is in the irreversible mode. 
In the bigger range of parameter space temperature does not facilitate 
energy transduction. By fine tuning the parameters one can obtain a regime 
in which temperature facilitates energy transduction. 
However, in this parameter space the attained value of 
efficiency is found to be in the subpercentage level. 

We also observe current reversals in the adiabatic limit by proper 
tuning of different parameters. These reversals are 
attributed to the complex dynamics of the  system. From 
our study of the nature of input energy and currents 
we conclude that there is no resonance phenomenon 
occurring in the system. The analysis of current and 
entropy production results show that the peak in current 
and entropy production do not coincide.  

It is worthwhile exploring whether the transport in these efficient 
ratchet is coherent or not. Noise induced currents are always accompanied by a
diffusive spread. If the diffusive spread of the particle is less than the
average distance (say, length of the period of the potential) travelled by the
particle in a given time then the transport is said to be coherent. 
This is quantified in terms of the so-called dimensionless Peclet 
number~\cite{pec}. The present work is concentrated mainly on 
optimizing the thermodynamic efficiency. One can as well optimize 
maximum work or have the best compromise between maximum work and efficiency. 
This can be done using known optimizing criterion~\cite{hernandez}. 
Studies in this regard is currently in progress.

 \end{document}